# Stark effect in a wedge-shaped quantum box.

Jorge-Alejandro Reyes-Esqueda[a,*], Carlos I. Mendoza[b], Marcelo del Castillo-Mussot[a], Gerardo J. Vázquez[a].

[a]Instituto de Física, UNAM, 04510, P. O. Box 20-364, 01000, México, D. F., Mexico.
[b]Instituto de Investigaciones en Materiales, UNAM, P. O. Box 70-360, 04510, México, D. F., Mexico.

**Abstract**

The effect of an external applied electric field on the electronic ground state energy of a quantum box with a geometry defined by a wedge is studied by carrying out a variational calculation. This geometry could be used as an approximation for a tip of a cantilever of an atomic force microscope. We study theoretically the Stark effect as function of the parameters of the wedge: its diameter, angular aperture and thickness; as well as function of the intensity of the external electric field applied along the axis of the wedge in both directions; pushing the carrier towards the wider or the narrower parts. A confining electronic effect, which is sharper as the wedge dimensions are smaller, is clearly observed for the first case. Besides, the sign of the Stark shift changes when the angular aperture is changed from small angles to angles $\theta > \pi$. For the opposite field, the electronic confinement for large diameters is very small and it is also observed that the Stark shift is almost independent with respect to the angular aperture.

[*] Corresponding author. Tel.: +52 55 5622-5184; Fax: +52 55 5616-1535.
*E-mail address*: reyes@fisica.unam.mx (J.-A. Reyes-Esqueda)

# 1. Introduction

In the last few years, the study of the properties of nanostructures has become very important in looking for new technological applications. Specially, the behavior of their electronic and optical properties in the presence of an external electric field has been widely studied, both theoretical and experimentally [1-8]. For example, in the case of semiconductor nanocrystals, the external field can be used to control electron transfer in quantum dot-based single electron devices and to control elementary qubit operations in such systems [9, 10]. Also, the electric field makes possible the mapping of the electronic wave functions of quantum dots in scanning-tunneling-microscopy experiments [11]. Furthermore, the mechanisms of carrier transport in quantum well structures and the capture process as function of an electric field, which may be due to a piezoelectric effect, have attracted much interest for the design and optimization of electronic and photonic devices [12].

On the other hand, from the standpoints of practical device application and fundamental science, new detection techniques using micro/nanomechanical beams and cantilevers have received increasing attention. Thus, research about the influence of quantum effects on the piezoresistance of two-dimensional heterostructures has been made [13]. Moreover, since cantilever based sensors have become a standard on micro- and nano-electromechanical systems (MEMS and NEMS) for detection of magnitudes with resolution in the pico-scale, there is recent work in developing electromechanical models for a transducer based on a lateral resonating cantilever [14], a feedback controlled nanocantilever device [15], and biosensors based on polymer microcantilevers [16]. Additionally, very recently, Qi Ye, et al, have beginning a large-scale fabrication of carbon nanotube probe tips for atomic force microscopy (AFM) imaging applications [17] and Liwei Chen, et al. have already made imaging of gold nanoclusters and biomolecules in ambient and fluid environments with this kind of AFM probes [18].

As said above, the study and understanding of electronic properties of systems on the nano-scale is very important for a wide range of possible applications. One example of this is the modeling of the tip and sample surfaces in scanning tunneling microscopy as confocal hyperboloids [19] and the consequent construction of their normal modes and Green functions by using prolate spheroidal coordinates [20]. Another example is the importance on electrostatic force gradient microscopy of the electrostatic forces between sharp tips and metallic and dielectric samples given the geometry of the tip [21]. Finally, it results that these electrostatic forces, and even Casimir force, have influence on the mechanical properties of cantilevers very dependent on the geometry of the cantilever tip [22]. Following this motivation and the experience obtained in the study of Stark effect with variational techniques in several geometries [7, 8, 23], we decided to study this effect in the tip of a typical cantilever of an atomic force microscope (AFM) using as a first approximation the geometry given by a wedge since, as a matter of fact, there exist commercial triangular-shaped silicon microcantilevers (Thermomicroscopes) [24]. Therefore, in the present work, as a first approximation to the AFM tip, we carry out a variational calculation to study the influence of an external applied electric field on the electronic ground-state energy of a wedge-shaped quantum box as function of the parameters of the wedge (diameter, angular aperture and thickness) and the intensity and direction of the electric field. This last is applied along the axis of the wedge in both directions, to the wider and to the narrower parts. For our calculations we suppose that the dielectric constant of the wedge is the same as the medium where it is embedded. In future works, we will consider variations of the dielectric constant of the wedge as well as other electric fields, calculations for excited states and the inclusion of impurities inside the tip.

In section 2 we present our calculations. Then, in section 3 we discuss our results, and finally we give our conclusions.

## 2. Calculation

In cylindrical coordinates, the Hamiltonian of a carrier in a quantum box defined by the geometry of a wedge with an electric field applied along its symmetry axis as shown in figure 1, is given by

$$H = \frac{p^2}{2m^*} \pm |e| F \rho \cos\theta + V_c(\rho, \theta, z), \qquad (1)$$

where $F$ is the electric field, $m^*$ and $|e|$ are the electron effective mass and charge, respectively, and $V_c$ is the confining potential, which vanishes inside the wedge and becomes infinite outside using the infinite well model. For electrons and $F>0$, the + sign corresponds to the electric field pointing to the positive direction of the axis $x$ and the - sign to the field in the opposite direction as shown in figure 1.

The energy levels of the quantum box in the absence of the field are given by

$$E_{n,m,l} = \frac{\hbar^2}{2m^*}\left[\left(\frac{\alpha_{m',n}}{d}\right)^2 + \left(\frac{l\pi}{L}\right)^2\right], \qquad (2)$$

where $\alpha_{m',n}$ is the $n$th zero of the Bessel function of order $m'$, $J_{m'}$; with $m' = m\frac{\pi}{\theta_0} = (2n_\theta+1)\frac{\pi}{\theta_0}$; $l = 2n_z+1$; $n = 1,2,3,\ldots$; $n_\theta, n_z = 0,1,2,\ldots$ and $d$, $L$ and $\theta_0$ are the dimensions of the wedge. For the ground state, $n=m=l=1$, this energy is given by

$$E_{1,1,1} = \frac{\hbar^2}{2m^*}\left[\left(\frac{\alpha_{m_0,1}}{d}\right)^2 + \left(\frac{\pi}{L}\right)^2\right], \qquad (3)$$

where $m_0 = \frac{\pi}{\theta_0}$, and the related wave function is

$$\Psi_0(\rho,\theta,z) = N_0 J_{m_0}\left(\frac{\alpha_{m_0,1}}{d}\rho\right)\cos\left(\frac{\pi}{\theta_0}\theta\right)\cos\left(\frac{\pi}{L}z\right), \qquad (4)$$

with $N_0$ the normalization constant. Evidently, $\Psi_0(\rho,\theta,z)=0$ outside the wedge.

Now, when the electric field is applied, we will use as our variational function this ground state wave function multiplied for an exponential factor, which accounts for the Stark effect, as follows:

$$\Psi(\rho,\theta,z) = \Psi_0(\rho,\theta,z)\exp(\mp\beta\rho\cos\theta), \qquad (5)$$

with $\beta$ the variational parameter that depends on the electric field. For electrons, the - sign corresponds to the electric field pointing to the positive direction of the $x$-axis and the + sign to the opposite field and vice versa for holes. We calculate the ground state energy of the system when there is an applied electric field by minimizing

$$E(\beta) = \langle\Psi|H|\Psi\rangle = \int_{-\frac{L}{2}}^{\frac{L}{2}}\int_{-\frac{\theta_0}{2}}^{\frac{\theta_0}{2}}\int_0^d \Psi H \Psi \rho\, d\rho\, d\theta\, dz \qquad (6)$$

with respect to $\beta$ and then, we define the Stark shift as

$$\Delta E = E(\beta) - E_{1,1,1}. \tag{7}$$

## 3. Results and Discussion

All the results will be presented in reduced atomic units (a.u.*), which correspond to a length unit of an effective Bohr radius, $a^* = \hbar^2\varepsilon/m^*e^2$, and an effective Rydberg, $R^* = m^*e^4/2\hbar^2\varepsilon^2$. The electric field is also given in atomic units as $F_0 = e/2\varepsilon a^{*2}$, with $\varepsilon=1$ as it was stated above.

It is worthwhile to discuss what happens to the ground state energy of the wedge as we vary its dimensions in the absence of applied electric field. In table 1, it is possible to observe that the dependence of the energy on the thickness $L/a^*$ is weak in the range from 1 to 100. This behavior is more notorious for small angles since the coefficient $\alpha_{m_0,1}$ in equation (3) sharply increases as $\theta_0$ decreases reaching values much larger than $\pi$, as it is shown in figure 2. This fact means that the first term in the right hand side in equation (3) is much greater than the second one, and then also its contribution to the energy. Moreover, we may observe from table 1 that for the smallest angle, $\pi/20$, the change of the energy is about two orders of magnitude as $d/a^*$ varies from 1 to 10.

Obviously, the Stark shift does not depend on the thickness $L$ since the confinement due to the electric field lies in the $x$ direction as it is shown in the Hamiltonian, equation (1). Hence, for simplicity and in order to better simulate the tip of the cantilever, we chose for the rest of the calculations to take the thickness as $L/a^*=1$.

For the electric field applied in the positive direction of the $x$-axis (i.e. to the wider part of the wedge), the electron is pushed to the tip of the wedge. Figure 3 shows the Stark shift as a function of radius for angles $\theta < \pi$. In this figure, it can be observed that the Stark shift of the system increases as the electric field increases and it is larger for smaller angles as it is also evident from figure 4. These two facts are a consequence of the quantum confinement of the electron in the tip since this confinement is responsible of an increment of the kinetic energy of the electron. On the other hand, for $\theta > \pi$, the geometry of the wedge changes to a *Packman-like* shape rather than a tip (as shown in the inset of figure 5 for $\theta_0=3\pi/2$). In this case, depending on the field, the Stark shift could be either positive or negative as it is also shown in figure 5.

In figure 6, $\Delta E$ is shown as a function of the electric field for various wedge radii. In the case $\theta_0=\pi/20$, it can be observed that for a given radius, the electric field enhances the energy of the electron since this is more strongly confined to the tip of the wedge. This fact is not true anymore for $\theta_0=3\pi/2$. Indeed, in this figure, it can be observed the change of the sign of $\Delta E$ as the radius increases for $\theta_0=3\pi/2$. For $d/a^*=1$, $E(\beta)$ is still larger than $E_{1,1,1}$, but for $d/a^*=5$ or 10 this is no longer true. The change of sign in $\Delta E$, for a given value of the electric field, is a direct consequence of the choice of the origin for the electric potential, $V_F=|e|Fx$, at $x=0$, as we will discuss later.

In figure 7, we plot the ground-state electronic density, $\Psi^2(x, y, z=0)$, for the wedge-shaped geometry with no applied electric field (panel a)) and with an applied electric field (panel b)). The effect of the field is to push the electron towards the tip of the wedge as shown by the maximum of $\Psi^2$, which represents the most probable position of the electron. On the other hand, for the *Packman-shaped* geometry, if there is no applied field, the situation is very similar to that of the wedge-shaped case in which $\Psi^2$ presents a maximum located at some point $x>0$, as shown in figure 8 a). When we apply a small electric field, the maximum of $\Psi^2$ is only slightly shifted towards the origin (figure 8 b)). However, if the electric field is strong enough, then the maximum shifts to a position with $x<0$ and $\Psi^2$ shows two peaks of the same height due to the shape (figure 8 c)). This height is the same because it is equally probable to find the electron located in any of the two sides (*jaws*) of the *Packman-shaped* geometry.

The change of sign in $\Delta E$ of figures 5 and 6 can be understood more clearly if we consider a one-dimensional analogy of the system. First, we discuss the wedge-shaped geometry. In absence of an electric field, the electron can be seen as a particle moving in a one-dimensional potential with a minimum near the position of the maximum of $\Psi^2$, as it is shown in figure 7 c). The horizontal line represents the ground-state energy, $E_{1,1,1}$, of this system. When an electric field is applied, the resulting potential in which the particle is moving can be approximated by the potential without electric field plus the potential of the applied electric field, $V_F=|e|Fx$. These potentials are shown in figure 7 d). As it can be seen in the figure, the resulting potential presents a minimum shifted towards the tip of the wedge, in accordance with the behavior of $\Psi^2$. It is also evident that the minimum is shallower than in the case in which there is no applied field, and therefore, its ground-state energy, $E(\beta)$, is larger than $E_{1,1,1}$, giving rise to a positive value of the Stark shift, $\Delta E>0$. The situation is very similar for the *Packman-shaped* geometry. In absence of applied field, the electron moves in a potential with a minimum located near the maximum of the electronic density, with a ground-state energy, $E_{1,1,1}$, as shown in figure 8 d). When a small electric field is applied, the resulting potential presents a shallower minimum shifted towards the origin but always with $x>0$. The corresponding ground-state energy, $E(\beta)$, is larger than $E_{1,1,1}$, giving again rise to a positive value of the Stark shift, $\Delta E>0$ (figure 8 e)). However, if the field is large enough, then the potential may present a deeper minimum shifted towards $x<0$, with a corresponding ground-state energy, $E(\beta)$, smaller than $E_{1,1,1}$. This gives rise to a negative value of the Stark shift, $\Delta E<0$, which is not present in the wedge-shaped geometry (figure 8 f)). As it results clear, this change in sign is only a consequence of choosing the origin of the electric potential at $x=0$, and it could be suppressed, for a given field, with a different choice. The important fact is that, the application of an electric field pointing to the wider part of the wedge pushes the electron to the opposite side, confining it either to the tip of the wedge or to its two tips for the *Packman-shaped* geometry.

For an electric field applied to the opposite direction (i.e. to the narrower part of the wedge) the electron is pushed to the wider side of the wedge and then the quantum confinement is smaller (if the field is not very strong) than in the case in which there is not applied field. Therefore, $\Delta E$ is always negative as it is shown in figure 9. It can be also observed that, given an angle and a radius, the magnitude of the Stark shift increases as the electric field is increased. However, the most important result in this case is that the Stark shift is very insensitive to the angular aperture of the wedge. For instance, when $F/F_0=10$, it can be observed in figure 9 that the Stark shift is almost the same for $\pi/20$, $\pi/15$ and $\pi/10$ and also for $\pi/2$, $\pi$ and $3\pi/2$. Moreover, we found that $E(\beta)$ tends to almost zero for small fields ($F/F_0=0.5$ and 1) as the radius is increased, while for $F/F_0=10$ it tends to a negative value, in an almost independent way of the angular aperture in both cases.

## 4. Conclusions

We have analyzed the effect of an applied electric field on the electronic ground-state energy of a quantum box with a geometry defined by a wedge. We have shown how the quantum confinement effect strongly depends on the geometry of the system and the direction of the electric field. When the field is applied directed to the wider part of the wedge, the electron is pushed to the opposite side, then being strongly confined either to the tip of the wedge or to its two tips for the *Packman-shaped* geometry, increasing in both cases its kinetic energy. However, due to the choice of the origin of the electric potential, for a particular value of the applied electric field, the Stark shift may even be negative. This change of sign may be avoided by changing adequately the origin of the electric potential. Finally, for the field applied in the opposite direction (to the tip of the wedge), there is no electronic confinement for the considered applied electric fields and, given that the electronic energy decreases, the Stark shift is always negative and is very insensitive to the angular aperture of the wedge. As a future work, we will consider the influence of the dielectric constant of the wedge, to incorporate more complicated applied electric fields, and calculate the energies of excited states as well as the inclusion of impurities in the system, in order to make a more realistic modeling of the tip of a typical cantilever of an atomic force microscope.

## 5. Acknowledgements

This work has been partially supported by grant IN-106201 (DGAPA, UNAM, México). J. A. Reyes-Esqueda also acknowledges illuminating discussions with Lorea Chaos Cador.

## 6. References

1. J. A. Brum, G. Bastard. Phys. Rev. B 31 (1985) 3893.
2. G. Bastard, E. E. Mendez, L. L. Chang, L. Esaki. Phys. Rev. B 28 (1983) 3241.
3. D. A. B. Miller, D. S. Chemla, T. C. Damen, A. C. Gossard, W. Wiegmann, T. H. Wood, C. A. Burrus. Phys. Rev. B 32 (1985) 1043.
4. Y. P. Feng, H. S. Tan, H. N. Spector. Superlattices Microstruc. 17 (1995) 267.
5. W. Sheng, J. P. Leburton. Phys. Rev. Lett. 88 (2002) 167401.
6. A. Stróżecka, W. Jaskólski, M. Zieliński, G. W. Bryant. Vacuum 74 (2004) 259.
7. G. J. Vázquez, M. Castillo-Mussot, C. I. Mendoza, H. N. Spector. Phys. Stat. Sol. (c) 1 (2004) S54. C. I. Mendoza, G. J. Vázquez, M. Castillo-Mussot, H. N. Spector. 1 (2004) S74.
8. V. A. Harotyunyan, K. S. Aramyan, H. Sh. Petrosyan, G. H. Demirjian. Physica E 24 (2004) 173.
9. P. Zhang, X-G. Zhao. J. Phys. Condensed Matter 13 (2001) 8389.
10. I. Shtrichman, C. Metzner, B. D. Gerardot, W. V. Schoenfeld, P. M. Petroff. Phys. Rev. B 65 (2002) 081303(R).
11. O. Millo, D. Katz, Y. W. Cao, U. Banin. Phys. Rev. Lett. 86 (2001) 5751.
12. L. Villegas-Lelovsky, G. González de la Cruz. J. Appl. Phys. 95 (2004) 4204.
13. H. Yamaguchi, S. Miyashita, Y. Hirayama. Physica E 24 (2004) 70. H. Yamaguchi, S. Miyashita, Y. Hirayama. Appl. Surface Sc. 237 (2004) 649.
14. J. Teva, G. Abadal, Z. J. Davis, J. Verd, X. Borrisé, A. Boisen, F. Pérez-Murano, N. Barniol. Ultramicroscopy 100 (2004) 225.
15. C. Ke, H. D. Espinosa. Appl. Phys. Lett. 85 (2004) 681.
16. X. R. Zhang, X. Xu. Appl. Phys. Lett. 85 (2004) 2423.
17. Q. Ye, A. M. Cassell, H. Liu, K-J. Chao, J. Han, M. Meyyappan. Nanoletters 4 (2004) 1301.
18. L. Chen, C. L. Cheung, P. D. Ashby and C. M. Lieber. Nanoletters 4 (2004) 1725.
19. G. Seine, R. Coratger, A. Carladous, F. Ajustron, R. Pechou, J. Beauvillain. Phys. Rev. B 60 (1999) 11045. E. Ley-Koo. Phys. Rev. B 65 (2002) 077401. G. Seine, R. Coratger, A. Carladous, F. Ajustron, R. Pechou, J. Beauvillain. Phys. Rev. B 65 (2002) 077402.

20. N. Aquino, E. Castaño, E. Ley-Koo, S. E. Ulloa. Rev. Mex. Fis. E 50 (2004) 54.
21. S. Gómez-Moñivas, L. S. Froufe-Pérez, A. J. Caamaño, J. J. Sáenz. Appl. Phys. Lett. 79 (2001) 4048. G. M. Sacha, J. J. Sáenz. Appl. Phys. Lett. 85 (2004) 2610.
22. A. A. Chumak, P. W. Milonni, G. P. Berman. Phys. Rev. B 70 (2004) 085407.
23. J. W. Brown, H. N. Spector. J. Appl. Phys. 59 (1986) 1179.
24. *For instance, these cantilevers were used by F. Tian, et al. to study the potential-induced surface stress of a solid electrode in an electrochemical cell*. F. Tian, J. H. Pei, D. L. Hedden, G. M. Brown, T. Thundat. Ultramicroscopy 100 (2004) 217.

## Figure Captions

**Figure 1.** The geometry of the wedge and the electric field applied along its symmetry axis in two directions: to the tip and to the wider part. This last corresponds to the positive direction of the axis *x*. $0 \leq \rho \leq d$; $-\theta_0/2 \leq \theta \leq \theta_0/2$; $-l/2 \leq z \leq l/2$.

**Figure 2.** It is shown the first zero of the correspondent Bessel function given the angular aperture, $\theta_0$, of the wedge.

**Figure 3.** The Stark shift of the electron's energy, $\Delta E$, is shown as a function of the wedge radius for two electric fields, which are applied directed to the positive direction of the axis *x*.

**Figure 4.** $\Delta E$ is shown as a function of the angular aperture $\theta_0$ for $d/a^*=5$ and $F/F_0=1$.

**Figure 5.** Similar to figure 3, but with angles $\theta \geq \pi$. The particular shape of the wedge in the case of $\theta=3\pi/2$ is shown in the inset.

**Figure 6.** $\Delta E$ is shown, for the two extreme angular apertures, as a function of the electric field directed to the positive direction of the axis *x* for various wedge radii.

**Figure 7.** The ground-state electronic density, $\Psi^2(x, y, z=0)$, for $\theta_0=\pi/20$ and $d/a^*=10$, is shown for $F/F_0=0$ (panel a)) and $F/F_0=10$ (panel b)). The analogue 1D potential is shown in panels c) and d). The Stark shift, $\Delta E$, is positive in this case.

**Figure 8.** The ground-state electronic density, $\Psi^2(x, y, z=0)$, for $\theta_0=3\pi/2$ and $d/a^*=2$, is shown for $F/F_0=0$ (panel a)), $F/F_0=0.5$ (panel b)) and $F/F_0=10$ (panel c)). The analogue 1D potential is shown in panels d), e) and f). The Stark shift, $\Delta E$, is positive or negative depending on the magnitude of the applied electric field.

**Figure 9**. Δ*E* is shown as a function of the wedge radius for various electric fields, which are applied directed to the tip of the wedge.

## Table captions

**<u>Table 1</u>**. The system electronic ground state energy when there is not applied field is shown as a function of its dimensions.

*J. A. Reyes-Esqueda, et al.*

Figure 1.

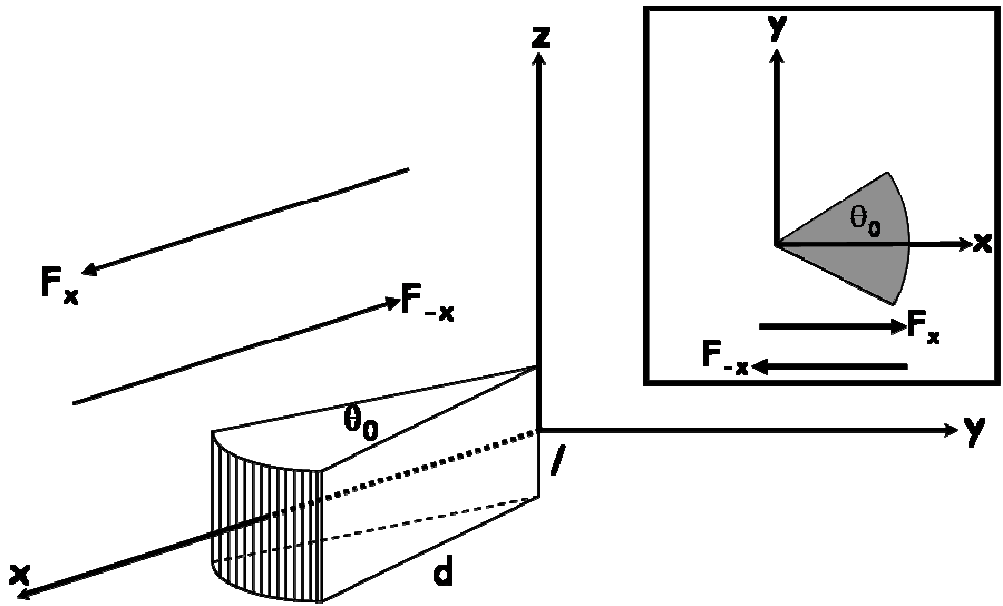

FIG. 1

Figure 2.

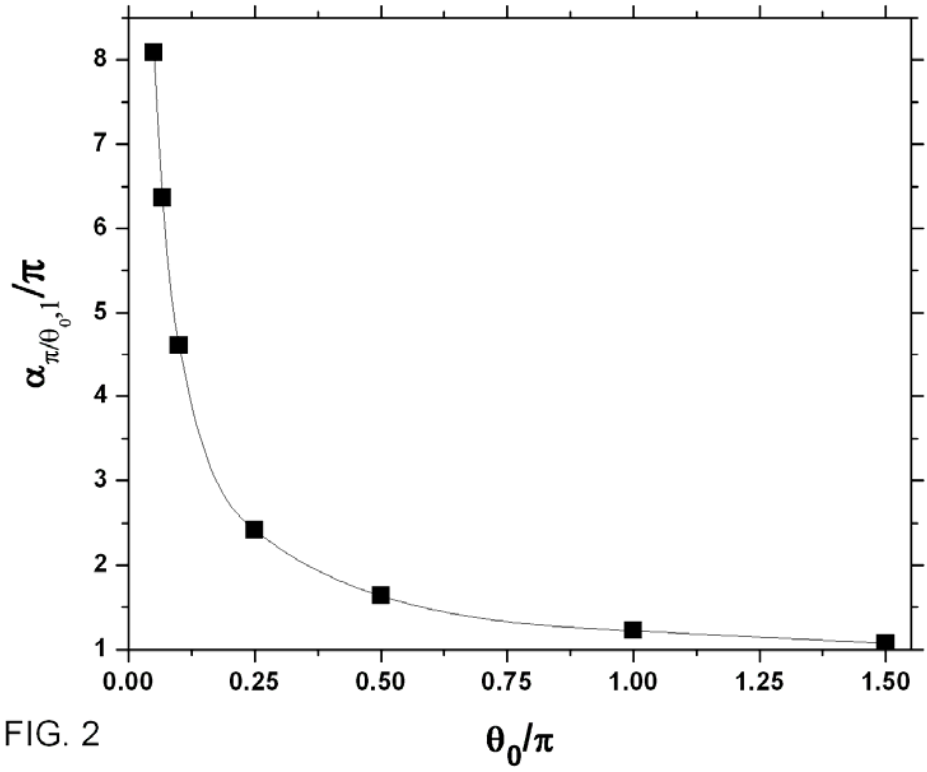

FIG. 2

Figure 3.

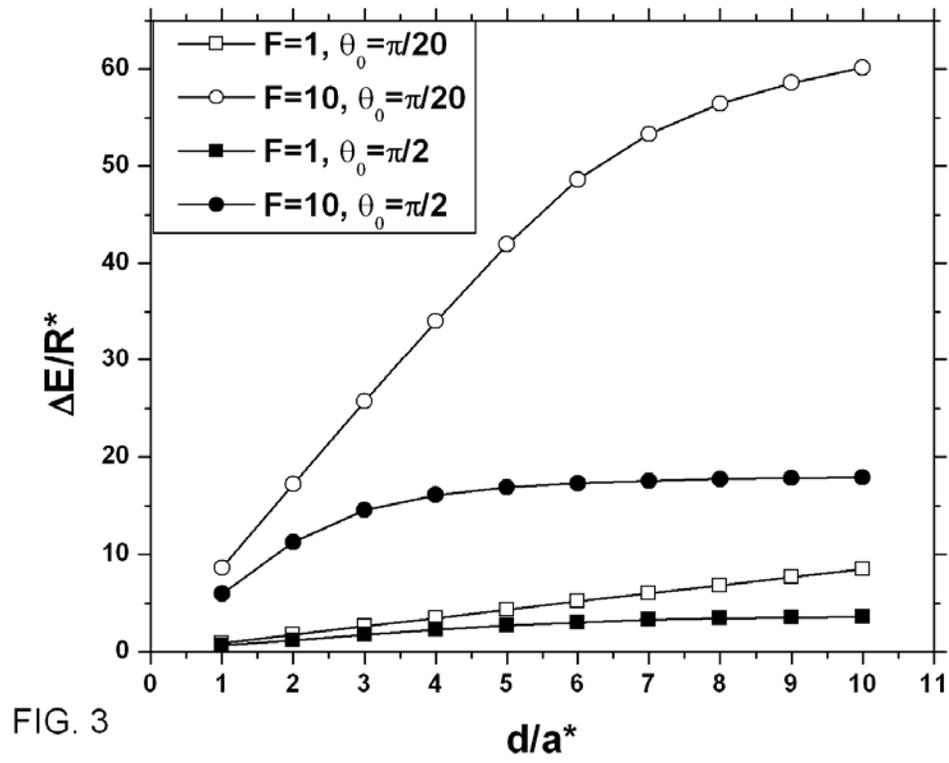

FIG. 3

*J. A. Reyes-Esqueda, et al.*

Figure 4.

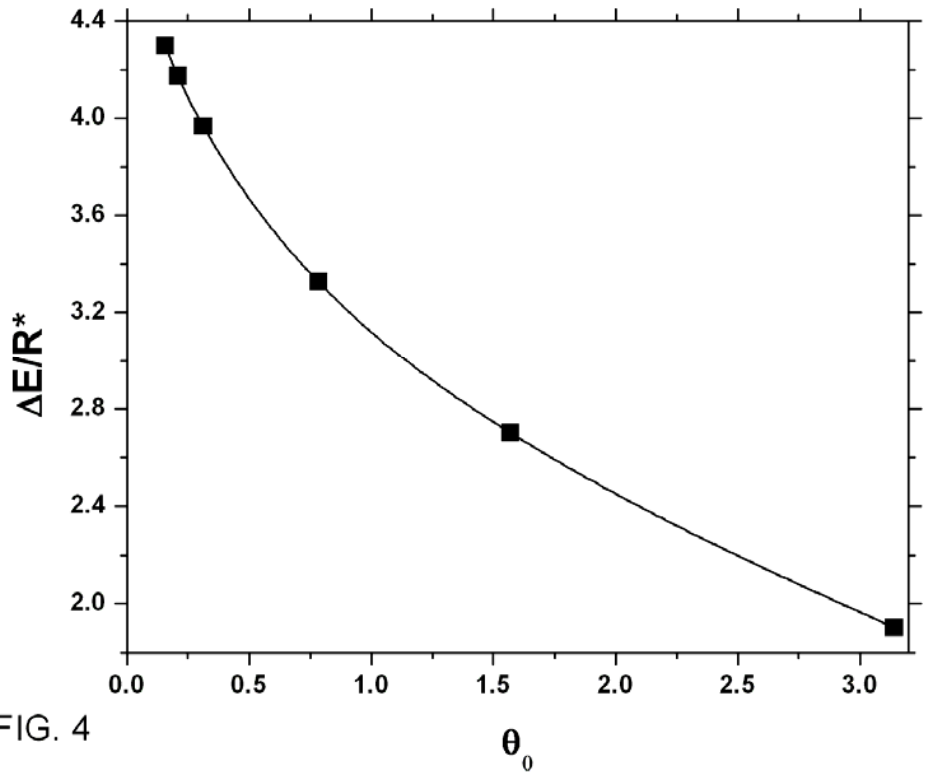

FIG. 4

*J. A. Reyes-Esqueda, et al.*

Figure 5.

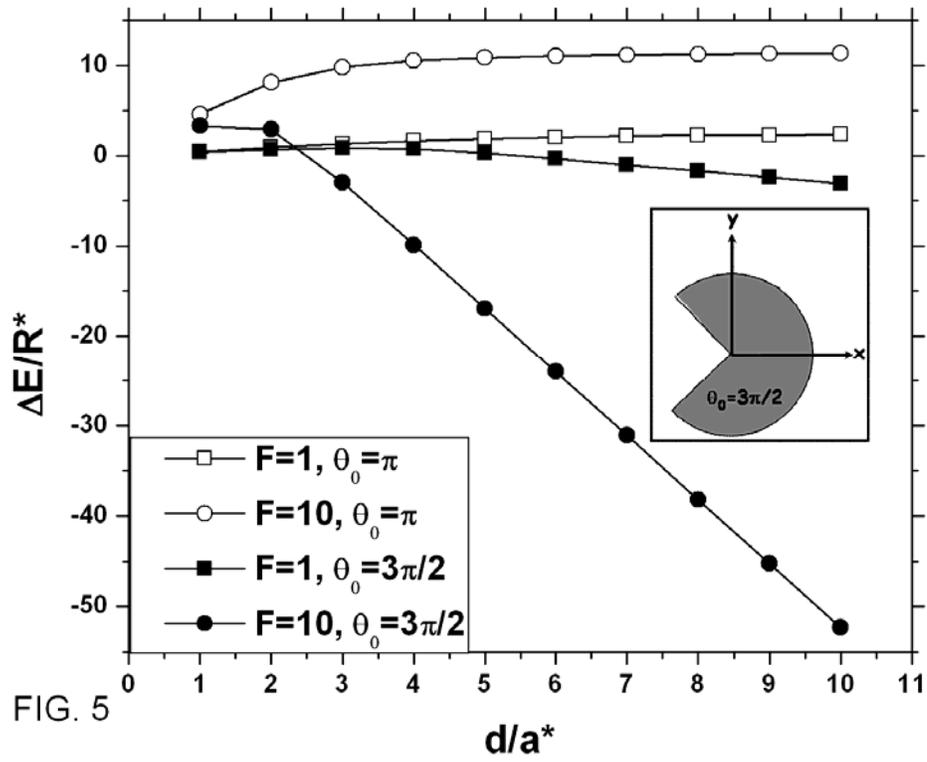

FIG. 5

*J. A. Reyes-Esqueda, et al.*

Figure 6.

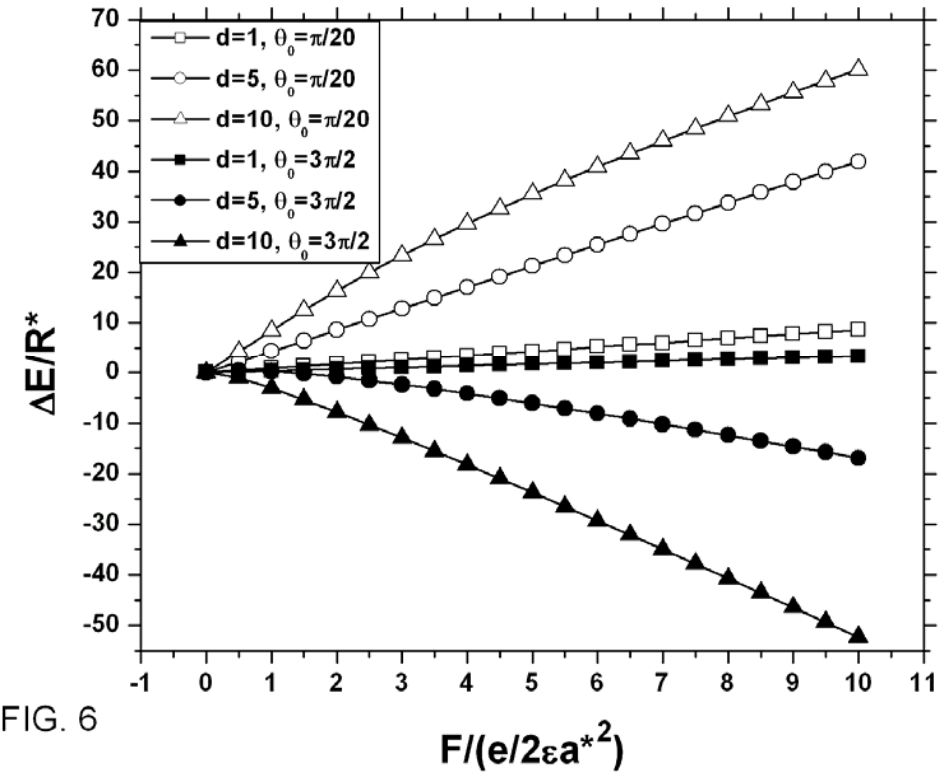

FIG. 6

*J. A. Reyes-Esqueda, et al.*

Figure 7.

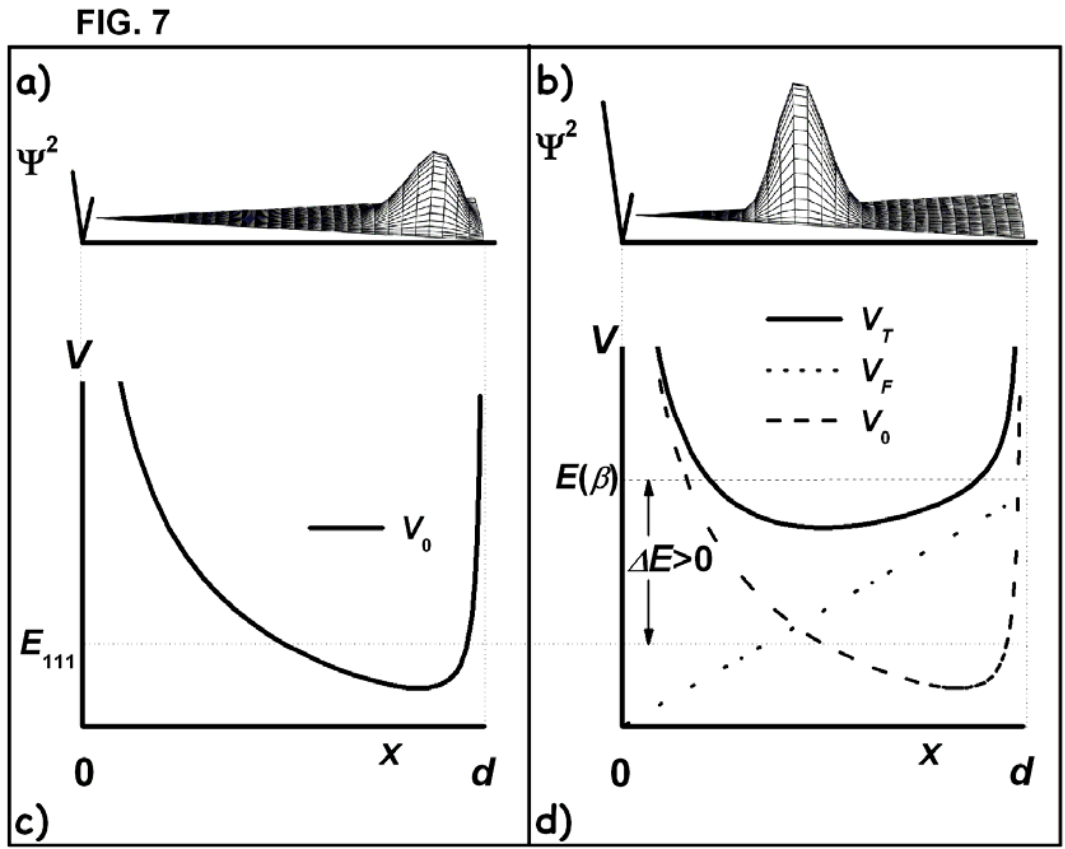

*J. A. Reyes-Esqueda, et al.*

Figure 8.

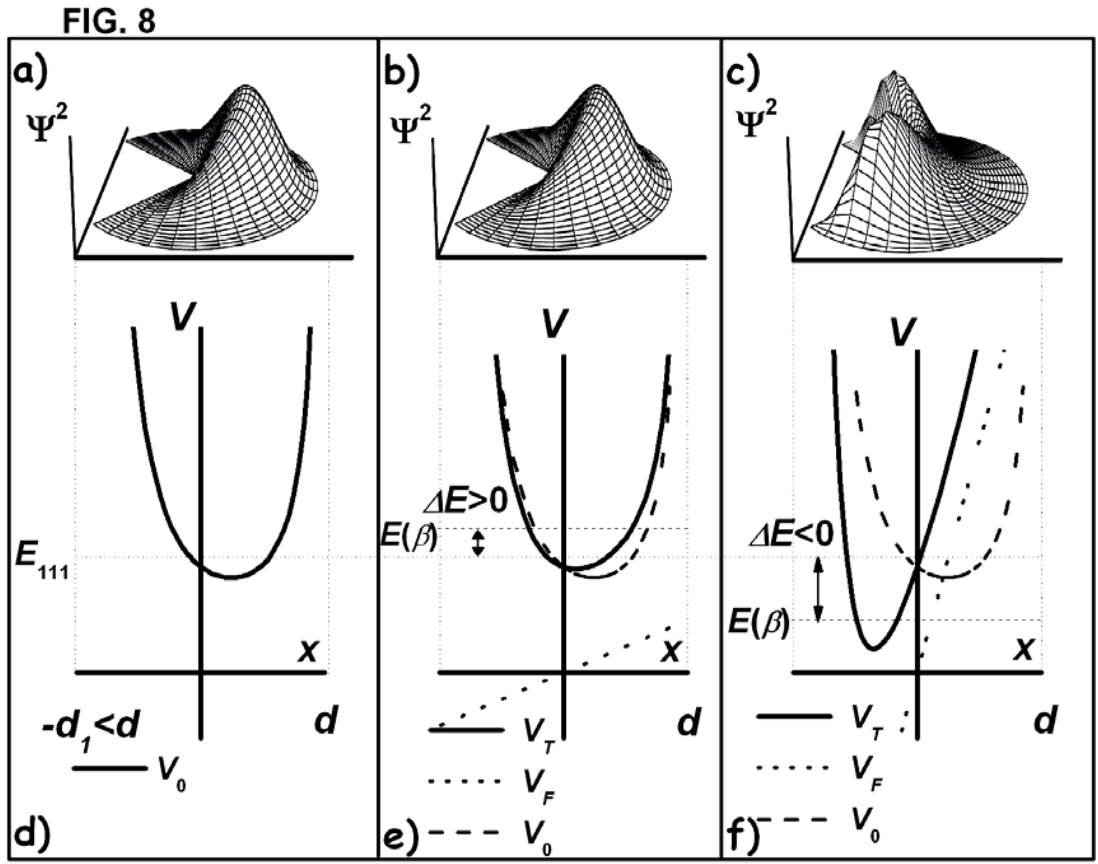

Figure 9.

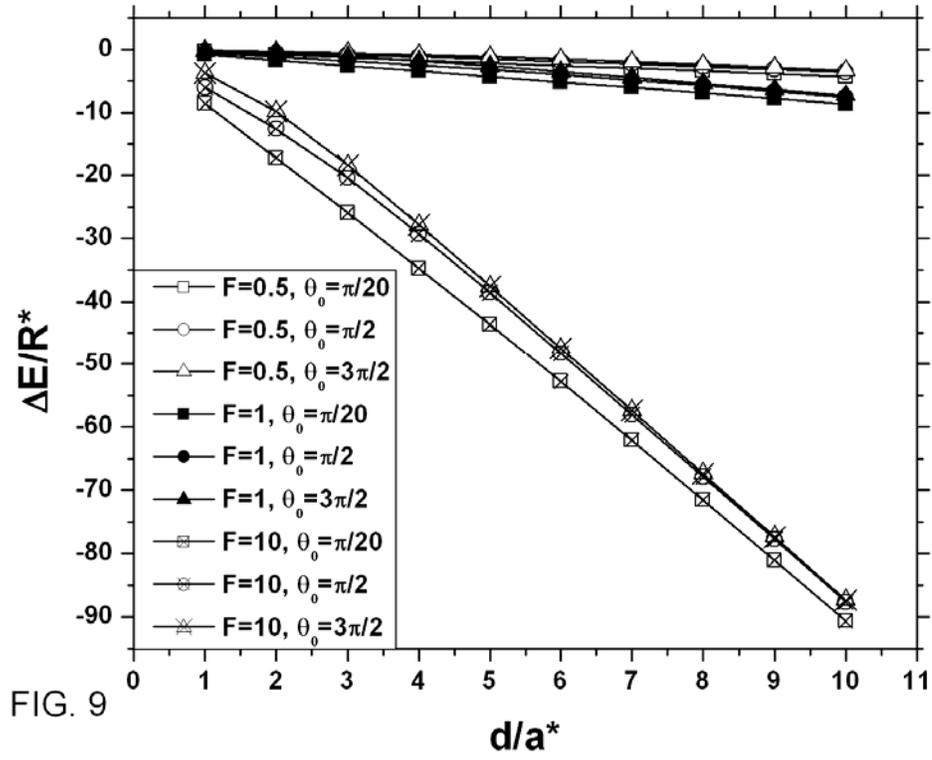

*J. A. Reyes-Esqueda, et al.*

Table 1.

| $d/a*$ | $L/a*$ | $E_{1,1,1}(\pi/20)$ | $E_{1,1,1}(\pi/10)$ | $E_{1,1,1}(\pi/2)$ | $E_{1,1,1}(\pi)$ | $E_{1,1,1}(3\pi/2)$ |
|---|---|---|---|---|---|---|
| 1 | 1 | 655.90 | 219.40 | 36.24 | 24.55 | 21.26 |
| 1 | 10 | 646.12 | 209.63 | 26.47 | 14.78 | 11.49 |
| 1 | 100 | 646.03 | 209.54 | 26.37 | 14.68 | 11.39 |
| 2 | 1 | 171.37 | 62.25 | 16.46 | 13.54 | 12.71 |
| 2 | 10 | 161.60 | 52.48 | 6.69 | 3.76 | 2.94 |
| 2 | 100 | 161.50 | 52.38 | 6.59 | 3.67 | 2.84 |
| 4 | 1 | 50.24 | 22.96 | 11.51 | 10.78 | 10.58 |
| 4 | 10 | 40.47 | 13.19 | 1.74 | 1.01 | 0.81 |
| 4 | 100 | 40.37 | 13.09 | 1.64 | 0.91 | 0.71 |
| 6 | 1 | 27.81 | 15.69 | 10.60 | 10.27 | 10.18 |
| 6 | 10 | 18.04 | 5.91 | 0.83 | 0.50 | 0.41 |
| 6 | 100 | 17.94 | 5.82 | 0.73 | 0.40 | 0.31 |
| 8 | 1 | 19.96 | 13.14 | 10.28 | 10.09 | 10.04 |
| 8 | 10 | 10.19 | 3.37 | 0.51 | 0.32 | 0.27 |
| 8 | 100 | 10.09 | 3.27 | 0.41 | 0.23 | 0.17 |
| 10 | 1 | 16.32 | 11.96 | 10.13 | 10.01 | 9.98 |
| 10 | 10 | 6.55 | 2.19 | 0.36 | 0.24 | 0.21 |
| 10 | 100 | 6.46 | 2.09 | 0.26 | 0.14 | 0.11 |